\documentclass[onecolumn,prd,amsmath,amsssymb,groupedaddress,12pt]{revtex4}
\usepackage{graphicx}
\usepackage{dcolumn}
\usepackage{bm}
\usepackage{amssymb}
\def\lsim{\mathrel{\mathstrut\smash{\ooalign{\raise2.5pt\hbox{$<$}\cr\lower2.5pt\hbox{$\sim$}}}}}
\def\gsim{\mathrel{\mathstrut\smash{\ooalign{\raise2.5pt\hbox{$>$}\cr\lower2.5pt\hbox{$\sim$}}}}}
\def\be{\begin{equation}}
\def\ee{\end{equation}}
\def\bea{\begin{eqnarray}}
\def\eea{\end{eqnarray}}

\def\cG{{\cal G}}
\def\cO{{\cal O}}
\def\cI{{\cal I}}

\def\cL{{\cal L}}
\def\cP{{\cal P}}
\def\l{\left}
\def\r{\right}

\begin{document}

\title{Fading Gravity and Self-Inflation}
\author{Justin Khoury}

\affiliation{Perimeter Institute for Theoretical Physics, 
31 Caroline St. N., Waterloo, ON, N2L 2Y5, Canada}

\begin{abstract}

We study the cosmology of a toy modified theory of gravity in which gravity shuts off at short distances, as in the fat graviton scenario of Sundrum.
In the weak-field limit, the theory is perturbatively local, ghost-free and unitary, although likely suffers from non-perturbative instabilities. 
We derive novel self-inflationary solutions from the vacuum equations of the theory, 
without invoking scalar fields or other forms of stress energy. The modified perturbation equation expressed in terms of the Newtonian potential closely resembles its counterpart for inflaton fluctuations. The resulting scalar spectrum is therefore slightly red, akin to the simplest scalar-driven inflationary models. 
A key difference, however, is that the gravitational wave spectrum is generically not scale invariant. In particular the tensor spectrum can have a blue tilt, a distinguishing feature from standard inflation. 
\end{abstract}

\maketitle

\section{Introduction}

What if gravity becomes {\it weaker} --- and perhaps even shuts off --- at short distances? This tantalizing possibility should have dramatic implications for early-universe cosmology. An immediate question is whether high energy inflation can still take place --- after all, the accelerated expansion of inflationary cosmology relies on the backreaction of vacuum energy on the geometry. But maybe shutting off gravity at high energy can obviate the need for inflation and offer interesting alternatives. If no gravity means no cosmic expansion, the universe could conceivably start out in a quiescent state, having plenty of time to thermalize and homogenize.

The most widely studied classical modifications of Einstein gravity, on the other hand, such as those arising in theories with extra dimensions, all have in common that gravity is {\it stronger} at short distances. This can be understood in a variety of ways. First, stronger gravity is seen in the Newtonian potential becoming steeper than $1/r$ above the compactification scale:
\be
V(r)\sim \frac{1}{r}\rightarrow \frac{1}{r^{D-3}}\,.
\ee
Stronger gravity can equivalently be understood at the level of the 4d effective theory: the various moduli describing the size and shape of extra dimensions mediate attractive yukawa forces, thereby enhancing gravity. Finally, stronger gravity leaves an imprint on the cosmological evolution.
For instance the modified Friedmann equation~\cite{deffayet,renjie} in the Randall-Sundrum brane-world scenario~\cite{rs2},  
\be
H^2 = \frac{8\pi G}{3}\rho \left(1+ \frac{\rho}{2\sigma}\right)\,,
\ee
implies faster expansion than in standard cosmology. But this is precisely what we expect from stronger gravity: for a spatially-flat universe to expand forever, as it must, the expansion rate must be higher to overcome the stronger gravitational pull. 

This paper considers instead the implications of weaker or fading gravity on early-universe dynamics.
The idea of ``asymptotic safety" for gravity~\cite{weinberg} has been exploited in various contexts~\cite{litim}.
It has been proposed to resolve black hole singularities, either through non-perturbative $\alpha'$ effects in string theory~\cite{tseytlin}, or in models inspired by the non-local form of the open string field theory action~\cite{witten,watireview,siegel}. 
Weaker gravity is featured in the recent ``fat graviton"~\cite{raman} proposal to address the cosmological constant problem.
An interesting search for fading gravity in brane-world constructions is described in~\cite{brane}.

We focus on a particular toy model of fading gravity, where the graviton propagator shuts off analytically in the UV: 
\be
\frac{1}{\Box} \rightarrow \frac{\cG\left(\Box L^2\right)}{\Box}\,.
\label{prop}
\ee
Here $\cG$ goes to zero on scales much smaller than $L$ and becomes trivial on large scales. (Alternatively one can suppress gravity with the addition of a massive or Proca spin-1 field that couples to matter with charge $m/M_{\rm Pl}$~\cite{scherk}. As in electromagnetism, a vector-mediated force is repulsive for like charges, quenching the effective gravity. See~\cite{will} for cosmological applications.) In order to avoid adding new poles in the propagator, which would invariably introduce ghosts, $\cG$ must be analytic, {\it e.g.},
\be
\cG\l(\Box L^2\r) = \exp\l(-\Box^2 L^4\r)\,. 
\label{fidG}
\ee
More generally, it can take the form $\exp\l(f\l(\Box L^2\r)\r)$, where $f$ is an arbitrary analytic function consistent with the asymptotic behavior of the form factor. This mechanism for shutting off gravity at short distances is inspired by the ``fat graviton" proposal~\cite{raman}. Analytic propagators of this form appear in Polchinski's formulation of the renormalization-group flow~\cite{pol}, and have been considered to regularize divergences in gauge theories~\cite{moffatwoodard}. In the cosmological context, modified propagators have been proposed to derive bouncing cosmologies~\cite{siegelcosmo}. With normalized kinetic term, the action is reminiscent of the tachyon sector of bosonic open string field theory as well as the scalar field action in $p$-adic string theory~\cite{padic}. 

The spatial and temporal nonlocality of our form factor leads to an action with infinitely many time derivatives, which therefore lacks 
a well-defined initial value formulation and likely suffers from non-perturbative Ostrogradskian instability~\cite{ostro,woodard}. 
Nevertheless the theory seems remarkably well-behaved in most regimes of interest for this work. 

In the weak field limit, for instance, the theory is perturbatively ghost-free, unitary and has a well defined initial value formulation. Indeed, despite arising from equations of motion with infinitely many time derivatives, all perturbative solutions are specified by a finite amount of initial data. This can be understood by the localization procedure~\cite{woodard}: any analytic non-local theory with local unperturbed lagrangian can be recast through field redefinitions as an action that is local in time and reproduces all perturbative solutions. Since any curved background is locally flat, in this sense the UV limit of our theory seems under control.

Even perturbatively, however, the theory is not without worries. It is well-known that analytic modifications to the propagator lead to acausality. At the classical level this is seen directly in perturbative solutions having support outside the past light-cone. In quantum field theory, the culprit for acausality is the essential singularity of~(\ref{fidG}) at infinity in the complex frequency plane which spoils the analytic properties of scattering amplitudes~\cite{schweber}. 
Be that as it may, at a philosophical level acausality at high energy is arguably a mild sin. It may even be a blessing for early-universe dynamics, in particular for
the homogeneity and horizon problems of standard big bang cosmology.

\subsection{Self-Inflation}

The modified Einstein equations, whose weak field limit describes a graviton with propagator~(\ref{prop}), are of the form
\be
\cG^{-1}\l(\Box L^2\r)G_{\alpha\beta} + \cO \l(R^2\r) = 8\pi G_{\rm N}T_{\alpha\beta}\,,
\label{sketch}
\ee
where the curvature-squared terms ensure the Bianchi identity. Focusing on the leading term, the form factor $\cG$ is recognized as an effective Newton's constant,
\be 
G^{\rm eff}_{\rm N} \l(\Box\r)= \cG\l(\Box L^2\r)G_{\rm N} \,,
\ee
which indeed vanishes at short distances. 

The corresponding modified Friedmann equation offers us a pleasant surprise: an accelerating universe without invoking scalar fields or any other form of stress energy. This novel inflationary solution instead arises from the vacuum equations of the modified gravity action --- a self-inflating solution. This is interesting phenomenologically since a lot is known about inflation with scalar fields: the required fine-tuning on scalar potentials, their generic predictions~\cite{kst,boyle}, etc. Our theory, however, is a genuine spin-2 modification and therefore cannot be rewritten as Einstein gravity plus scalars. Thus the resulting accelerating solution is different than scalar-driven inflation and, as we will see, leads to distinguishing observational signatures. 

To understand how self-inflation is possible, a key observation is that de Sitter space, or any Einstein space for that matter, identically satisfies $\Box G_{\alpha\beta}^{\rm dS} = 0$. Thus the action of $\cG^{-1}$ on the Einstein tensor is trivial, $\cG^{-1}\l(\Box L^2\r)G_{\alpha\beta}^{\rm dS}  = G_{\alpha\beta}^{\rm dS}$, no matter how small the dS radius is compared to $L$. 

Now consider a geometry which is nearly de Sitter, but slowly-evolving, in the sense that $|\dot{H}/H^2|\ll 1$. That is, our ansatz is $R_{\alpha\beta} = 3H^2g_{\alpha\beta} + r_{\alpha\beta}$, with $r$ a small correction of order $\dot{H}$. Since $r_{\alpha\beta}$ is slowly-changing, we can neglect its time-derivatives, which amounts to dropping terms of order $\ddot{H}$, $\dot{H}^2$, etc. Schematically, we find
\be
\cG^{-1}\l(\Box L^2\r)r_{\alpha\beta}\sim \cG^{-1}\l(8H^2L^2\r)\dot{H}\,.
\ee
Thus, although $r_{\alpha\beta}$ constitutes a small deviation from exact de Sitter, its contribution to the equation of motion is ``boosted" by a factor of $\cG^{-1}\l(8H^2L^2\r) \gg 1$ for $HL\gsim 1$. 

Our solution therefore can be understood as arising from a compensation between the de Sitter $H^2$ term and the ``boosted" $\dot{H}$ correction:
\be
3H^2 + 8\cG^{-1}\l(8H^2L^2\r)\dot{H}\approx 0 \,,
\ee
that is,
\be
\frac{\dot{H}}{H^2} \approx -\frac{8}{3}\cG\l(8H^2 L^2\r)\,.
\label{eom1}
\ee
The factor of $\cG\l(8H^2 L^2\r)$ on the right-hand side is indeed small provided that $HL\gsim 1$, justifying the slow-evolution approximation. The latter
can be checked explicitly by taking derivatives of~(\ref{eom1}) and noting that $\dot{H}^2$, $\ddot{H}$ etc. are down by extra powers of $\cG$. While this self-inflating solution relies on the higher derivative nature of the theory, nevertheless the approximate equation of motion~(\ref{eom1}) from which it derives is second order in the scale factor, just as in Einstein gravity. In this sense the dependence on the higher derivative structure is the most minimal possible. 

Intuitively, self-inflation can be understood as follows~\footnote{We thank L.~Sorbo for suggesting this compelling interpretation.}. The fading of gravity arises from self-interactions of the graviton which become relevant on scales smaller than $L$. These self-interactions in turn give rise to the curvature-squared terms in~(\ref{sketch}), which manifestly act as a source for gravity. However, thanks to the filter function this source inevitably appears to gravity as approximately homogeneous in space and time on scales smaller than $L$ --- it therefore plays the role of an effective vacuum energy, causing the universe to inflate.

Surprisingly, de Sitter space is an approximate solution of increasing accuracy for larger values of $HL$, and generic higher-curvature corrections to the effective action also become less and less relevant in this limit.  The latter either involve derivatives acting on curvature tensors, which means they are subleading in $\dot{H}$; or they are higher order in curvature, in which case they simply ``renormalize" the form factor $\cG$. This is in stark contrast with the usual situation in effective field theory where corrections to the effective action become increasingly relevant above the scale of new physics. This highly desirable property of our accelerating solution traces back to pure de Sitter being oblivious to the form factor.

The inflationary era terminates when $\dot{H}\sim H^2$; that is, when $H\sim L^{-1}$. The details of the transition to decelerated expansion have yet to be worked out through numerical analysis, which we leave for future work. Nevertheless, we know by construction that the cosmological equation reverts to the usual Friedmann law for $HL\ll 1$. Moreover, when $\dot{H}\sim H^2$ the universe reheats to a temperature of order $T\sim L^{-1}$ through gravitational particle production~\cite{ford,others}. Since there is no inflaton field that couples directly to matter, this is the only way to reheat. Therefore, we conjecture that the universe exits the inflationary phase smoothly and enters a radiation-dominated epoch governed by the usual Friedmann equation. 

\subsection{Perturbation Spectrum}

Our accelerating solution can be understood in a different way, which is useful for studying the generation of density perturbations. Anticipating an approximate de Sitter solution, consider the metric ansatz
\be
ds^2 = - e^{2\phi(t)} dt^2 + e^{-2\phi(t)}  e^{2\bar{H}t}dx^2\,,
\ee
where $\bar{H}$ is constant and $\phi$ is assumed small. At linear order, the latter is just the Newtonian potential. Our modified gravity equations can then be cast as an equation of motion for $\phi$:
\be
3\bar{H}\dot{\phi} \approx 8\bar{H}^2\cG \l(8\bar{H}^2L^2\r)\,.
\ee
In analogy with the slow-roll equation in standard inflation, $3H\dot{\varphi}\approx -V_{,\varphi}$, we see that the Newtonian potential plays the role of an inflaton, setting a clock for the cosmological evolution. (To make contact with~(\ref{eom1}), note that $\dot{H} = -\bar{H}\dot{\phi}\approx -H\dot{\phi}$.)

Density perturbations therefore originate from fluctuations in the Newtonian potential, $\Phi(\vec{x},t)$. In the limit where modes are well within the horizon, the geometry is approximately flat, and the equation for $\Phi_k$ reduces to
\be
\cG^{-1}\l(k^2 L^2\r) k^2 \Phi_k \approx 0\,.
\ee
Since $\cG$ is analytic, there are no new homogeneous solutions other than the usual plane waves in flat space, $\Phi_k\sim e^{-ikt}/\sqrt{2k}$. Meanwhile, in the opposite limit where modes are well outside the horizon, $\Phi_k$ satisfies
\be
3H\dot{\Phi}_k \sim H^2\cG \l(8H^2L^2\r)\Phi_k\,,
\ee
where we have dropped the bars for simplicity. Since $\cG \l(8H^2L^2\r)\ll 1$, an approximate solution is $\Phi_k\approx {\rm const.}$ These two extreme limits coincide with the corresponding limiting equations satisfied by inflaton perturbations in usual slow-roll inflation. By analogy, it follows that $\Phi$ acquires a scale-invariant spectrum with amplitude $\Delta\Phi_k \sim H/M_{\rm Pl}$, corresponding to a curvature perturbation
\be
\zeta \sim \frac{H}{\epsilon}\,,
\label{amp}
\ee
where $\epsilon \equiv -\dot{H}/H^2$ is a generalization of the usual $\epsilon$ parameter of slow-roll inflation.

Compared to the standard slow-roll result, $\zeta_{\rm slow-roll} \sim H/\epsilon^{1/2}$, the amplitude of perturbations in our case is enhanced by a factor of $\epsilon^{-1/2}$, which traces back to the difference in the generation mechanism. In slow-roll inflation, perturbations in the inflaton field have amplitude $\langle\delta\varphi\rangle\sim H$ and are related to Newtonian potential through $\delta\varphi = -\sqrt{2/\epsilon}\l(\dot{\Phi}/H + \Phi\r)$. In our case, the Newtonian potential itself acquires $\langle\Phi\rangle\sim H$, and this $\sqrt{\epsilon}$ mismatch trickles down to the final answer.

The scalar power spectrum is generically slightly {\it red} in fading gravity. Indeed, $\epsilon$ is proportional to $\cG(8H^2L^2)$ and therefore increases with time, while $H$ decreases with time. It follows from~(\ref{amp}) that modes which exit the horizon earlier have a larger amplitude, corresponding to a red spectrum. More precisely, for the choice of filter~(\ref{fidG}), we find $n_s\approx 0.96$ on the observable range of scales, in good agreement with current data~\cite{wmap3}. A blue scalar spectral tilt is also possible if one is willing to consider a non-monotonic $\cG$.

A key difference with standard inflation lies in is the spectrum of primordial gravitational waves: it is generically not scale invariant. This is not surprising since
our fading gravity model is a spin-2 modification, and should therefore strongly affect the tensor modes. While it is nevertheless possible to achieve a scale-invariant tensor spectrum by judiciously choosing the form factor, in general this leads to a gravity wave amplitude on large scales well below the projected sensitivity of near future experiments. This follows from the fact that the tensor-to-scalar ratio, $r$, is proportional to $\epsilon^2$ in our model, as seen from~(\ref{amp}), as opposed to $\epsilon$ in standard inflation. Thus $r$ is at best $10^{-4}-10^{-3}$ in fading gravity. Hence, a significant level of gravity waves on large scales requires a relatively strong red tilt, which, if observed, would distinguish our model from scalar-driven inflation. The fiducial form factor~(\ref{fidG}), for instance, yields a strong red tilt.

Another interesting feature is that the gravity wave spectrum is allowed to have instead a slight blue spectral tilt. Thus, if futuristic microwave background polarization or gravity wave detectors ever reached the $10^{-4}-10^{-3}$ sensitivity level for $r$, observing such a small blue tilt would strongly hint at fading gravity inflation. This is because a slightly red tensor spectrum is a robust prediction of scalar-driven inflation: since $\dot{H}= -\dot{\phi}^2/2 \leq 0$, independent of the potential, the gravity wave amplitude $\sim H/M_{\rm Pl}$ is a decreasing function of time and hence a decreasing function of scale. Blue gravity waves also appear in the ekpyrotic scenario~\cite{ek1,ek2,ek3}, but with such a strong spectral tilt that the large-scale amplitude is orders of magnitude below observable levels. A small blue tilt for tensors has also been obtained recently from thermal fluctuations of strings near the Hagerdorn temperature~\cite{nayeri}, as well as in super-inflationary models~\cite{paolo} based on ghost condensation~\cite{nimaghost}.

Perturbation analysis about the self-accelerating solution reveals a modified graviton propagator with 3 poles, one of which is ghost-like with a mass of order of the Hubble scale, indicating an instability of the solution. Intriguingly the ghost decouples in the limit $HL\rightarrow\infty$ in which de Sitter becomes an exact solution. To make the decay rate of the vacuum acceptably small, we must as usual invoke explicit Lorentz symmetry breaking at some high scale $\Lambda_{\rm LI}$~\cite{cline}. (A Lorentz-invariant cut-off yields a divergent rate.) An explicit calculation for the form factor~(\ref{fidG}) gives $\Lambda_{\rm LI}\lsim 10^{15}$~GeV. 

Once inflation ends and we enter the regime $HL\ll 1$, the ghost further decouples and becomes a non-perturbative excitation. Indeed, by construction our theory has only two perturbative propagating degrees of freedom about flat space --- the usual modes of the massless graviton. This decoupling is starkly different from theories of the form $R+c_1R^2+c_2R_{\alpha\beta}R^{\alpha\beta}$~\cite{stelle} and $F\l(R,R_{\alpha\beta}R^{\alpha\beta},R_{\alpha\beta\gamma\delta}R^{\alpha\beta\gamma\delta}\r)$~\cite{chiba,felice} where the Weyl ghost persists all the way to the UV cutoff.

The paper is organized as follows. We begin with a detailed exposition of the model in Sec.~\ref{model}. Given the complexity of the resulting equations of motion, we present three alternative derivations of the self-inflating solution. Exit from self-inflation and reheating are the subject of Sec.~\ref{exit}. In Sec.~\ref{stability} we argue that our self-inflating solution is stable under generic corrections to the effective action. Section~\ref{pert} deals with the generation of scalar density perturbations, their amplitude, and their spectral tilt. Tensor modes are discussed in Sec.~\ref{GW}. Section~\ref{ostro} is devoted to a general discussion of pathologies of higher time derivative theories, in particular the Ostrogradski instability, acausality, and the breakdown of the initial value problem. We conclude in Sec.~\ref{conclu} with a brief discussion of future research avenues and open problems.

\section{Fading Gravity Model}  \label{model}

The cut-off graviton propagator~(\ref{prop}) derives from the relativistic and covariant action
\be
S = \frac{M_{\rm Pl}^2}{2}\int d^4x \sqrt{-g}\left\{ R + G^{\alpha\beta}\left(\frac{\cG^{-1}\l(\Box L^2\r) - 1}{\Box}\right)R_{\alpha\beta}\right\}\,,
\label{action1}
\ee
where $\Box\equiv \nabla^\alpha\nabla_\alpha$ is the covariant d'Alambertian. Indeed, in the weak field limit, $g_{\alpha\beta} = \eta_{\alpha\beta} + h_{\alpha\beta}$, the Ricci tensor in Coulomb gauge is given by $R_{\alpha\beta} = -\Box h_{\alpha\beta}/2 + \cO(h^2)$, and~(\ref{action1}) reduces to
\be
S_{\rm weak-field} = \frac{M_{\rm Pl}^2}{8}\int d^4x \left(h^{\alpha\beta}-\frac{1}{2}\eta^{\alpha\beta}h\right)\cG^{-1}\l(\Box L^2\r)\Box h_{\alpha\beta}\,,
\ee
where the propagator displays the correct massless spin-2 tensor structure but is cut-off in the ultraviolet. 

In order to avoid new poles, which would invariably include ghosts, the filter function $\cG$ must be of the form $\exp\l(f\l(\Box L^2\r)\r)$, where $f$ is analytic. For concreteness, for most of the paper we will focus on the fiducial choice~(\ref{fidG}), $\cG\l(\Box L^2\r) = \exp\l(-\Box^2 L^4\r)$, where the even power of $\Box$ ensures that filtering occurs independently of whether the given mode has time-like or space-like momentum. 

For this fiducial choice, the form factor in~(\ref{action1}) can be expanded as $\l(\cG^{-1}-1\r)/\Box\approx \Box L^4 + \ldots$ and therefore does not include $1/\Box$ operators. It does, however, display an infinite number of time derivatives, which is {\it a priori} worrisome since temporally non-local theories generically suffer from instability and are bereft of an initial value problem. As we will argue in Sec.~\ref{ostro}, however, things seem under control for all regimes of interest here, including the self-inflating solution discussed below. 

Putting non-locality aside, the action~(\ref{action1}) nevertheless appears fine-tuned from the point of view of effective field theory. 
In general one expects $\cO\l(R^3\r)$ terms, more general derivative structure, and so forth. We will come back to this issue in Sec.~\ref{stability} and show that our self-accelerating solution is in fact robust under the inclusion of a wide class of generic corrections to the effective action.

\subsection{Fading Einstein's Equations}

The derivation of the equation of motion from~(\ref{action1}) is straightforward, modulo a few subtleties which are worth mentioning here.
First, we must be careful in varying the form factor since $\Box$ and $\delta\Box$ do not commute. For the fiducial choice~(\ref{fidG}) the final expression is
\be
\delta\left(\frac{\cG^{-1} -1}{\Box}\right) = -\frac{1}{\Box}\delta\Box\left(\frac{\cG^{-1}-1}{\Box}\right)-L^4\int_0^1ds\frac{\cG^{-s}}{\Box}\l(\delta\Box\cdot\Box+\Box\cdot\delta\Box\r)\cG^{s-1}\,,
\ee
where $\delta\Box$ implicitly takes into account the rank of the tensor on which it acts. Second, when varying the Ricci tensor,
\be
\delta R_{\alpha\beta} = -\frac{1}{2}\Box\delta g_{\alpha\beta} + \nabla^\gamma\nabla_{(\alpha}\delta  g_{\beta)\gamma}-\frac{1}{2}\nabla_{\alpha}\nabla_{\beta}\delta g\,,
\ee
one must keep in mind that while the last two terms yield a total derivative for the Einstein-Hilbert term, they give a non-trivial contribution
in the variation of the curvature-squared term.

At the end of the day the modified vacuum Einstein's equations are given by:
\bea
\nonumber
& & \cG^{-1}\l(\Box L^2\r)G_{\alpha\beta}  =  \cI^{(1)}_{\alpha\beta} + \int_0^1 ds\left(\cI^{(2)}_{\alpha\beta} +\cI^{(3)}_{\alpha\beta}\right)-\frac{1}{2}g_{\alpha\beta}R^{\gamma\delta} \left(\frac{\cG^{-1} -1}{\Box}\right)R_{\gamma\delta} \\
& & + \left(R_{\alpha\beta} -\frac{1}{4}g_{\alpha\beta} R\right)\left(\frac{\cG^{-1} -1}{\Box}\right)R + \left(2\delta_{(\alpha}^{\;\gamma}\delta_{\beta)}^{\;\delta}-g_{\alpha\beta}g^{\gamma\delta}\right)\nabla^\rho\nabla_\gamma\left(\frac{\cG^{-1} -1}{\Box}\right)G_{\rho\delta} \,,
\label{modeom}
\eea
where the $\cI$'s are defined below. The left hand side is just as in Einstein gravity, except for the coefficient which is recognized as an effective, scale-dependent Newton's constant, $G^{\rm eff}_{\rm N} (\Box)\sim \cG\l(\Box L^2\r)$, enforcing weaker gravity at short distances. 

The terms quadratic in curvature on the right hand side are necessary to maintain the Bianchi identity, $\nabla^\alpha \delta S/\delta g^{\alpha\beta}=0$, guaranteed by diffeomorphism invariance of the action. Indeed, taking the divergence of the left hand side and commuting the covariant derivative through $\cG^{-1}$ yields a host of terms quadratic in curvature, which precisely cancel the contribution from the right hand side.

The $\cI$ terms are of the form
\bea
\nonumber
\cI_{\alpha\beta} &\equiv & 2\nabla_{\alpha} B^{\gamma\delta}\nabla_{[\gamma} A_{\beta]\delta} +  \frac{1}{2}g_{\alpha\beta}\nabla_\kappa \left(B^{\gamma\delta}\nabla^\kappa A_{\gamma\delta} \right) +2\l(\nabla^\gamma\nabla_{\beta}B_{[\gamma}^{\;\;\delta}\r)A_{\alpha]\delta}+2\l(\nabla^\gamma\nabla_{\beta}A_{[\gamma}^{\;\;\delta}\r)B_{\alpha]\delta}\\
&+& 2\nabla_\alpha A_{\delta[\gamma}\nabla^\gamma B_{\beta]}^{\;\;\delta}-\nabla_\alpha B_{\beta}^{\;\;\gamma}\nabla^\delta A_{\gamma\delta}-B_{\alpha}^{\;\;\gamma}\Box A_{\beta\gamma} + A_{\gamma\alpha}\Box B_{\beta}^{\;\;\gamma}\,,
\label{I}
\eea
where symmetrization under $(\alpha,\beta)$ indices is implicit everywhere. The tensors $A$ and $B$ to be substituted in~(\ref{I}) are in turn given by
\bea
\nonumber
& & A_{\alpha\beta}^{(1)}\equiv \left(\frac{\cG^{-1} -1}{\Box}\right)R_{\alpha\beta}\;; \qquad B_{\alpha\beta}^{(1)}\equiv \frac{1}{\Box}G_{\alpha\beta}\;; \\
\nonumber
& & A_{\alpha\beta}^{(2)}\equiv L^4\Box\cG^{s-1}R_{\alpha\beta}\;; \qquad \;\;\;\;\; B_{\alpha\beta}^{(2)}\equiv \frac{\cG^{-s}}{\Box}G_{\alpha\beta}\;; \\
& & A_{\alpha\beta}^{(3)} \equiv L^2 \cG^{s-1}R_{\alpha\beta}\;;  \qquad \;\;\;\;\;\;\;  B_{\alpha\beta}^{(2)}\equiv \cG^{-s}G_{\alpha\beta}\,.
\eea
 
Finally, since we are interested in the cosmological evolution, we can focus on the trace of~(\ref{modeom}):
\bea
\nonumber
0 &=& -\cG^{-1}\l(\Box L^2\r)R - 2R^{\alpha\beta}\left(\frac{\cG^{-1} -1}{\Box}\right)R_{\alpha\beta} + 2\nabla^\alpha\nabla^\beta\left(\frac{\cG^{-1} -1}{\Box}\right)G_{\alpha\beta}  \\
&+& \cI^{(1)}+ \int_0^1 ds\left(\cI^{(2)}+\cI^{(3)}\right)\,.
\label{trace}
\eea
One might worry that taking the trace can yield solutions which implicitly hinge on a non-vanishing trace-free stress tensor, such as radiation. This is of no consequence here since the solution of interest is self-accelerating --- a radiation component would rapidly redshift away.

Given the level of complexity of the above equations of motion, it is not {\it a priori} obvious how to extract approximate solutions. In the remainder of the section we therefore provide three alternative ways of deriving our self-accelerating background: {\it i)} by direct substitution of the cosmological ansatz in a slow-evolution approximation (Sec.~\ref{direct}); {\it ii)} by assuming that the Ricci tensor is that of an Einstein space plus a small correction (Sec.~\ref{altern1}); {\it iii)} by substituting a metric ansatz which is de Sitter plus perturbation (Sec.~\ref{altern2}).

\subsection{Slow-Evolution Approximation} \label{direct}

Our key approximation assumes that the universe has slowly-evolving Hubble parameter: $|\dot{H}/H^2|\ll 1$, allowing us to neglect all terms
of order $\dot{H}^2$, $\ddot{H}$, etc. This greatly simplifies the equations of motion. 

Starting with the leading term in~(\ref{trace}), we have
\be
\cG^{-1}R = \left(1+\Box^2L^4+\ldots\right)R \approx 12H^2 + 6\dot{H} \,,
\label{term1}
\ee
where the last step follows from $\Box^2R\sim \cO\l(\dot{H}^2,\ddot{H}\r)$. Similarly, the second term is straightforward to compute:
\be
- 2R^{\alpha\beta}\left(\frac{\cG^{-1} -1}{\Box}\right)R_{\alpha\beta} \approx \frac{27}{4} \l(8H^2L^2\r)^2\dot{H}\,.
\label{term2}
\ee
The third term, however, is more subtle. Taylor-expanding the form factor and computing each term explicitly, we note the following structure: $2\nabla^\alpha\nabla^\beta\l(\Box L^2\r)^j L^2G_{\alpha\beta} \approx -(9/2)\l(8H^2L^2\r)^{j+1}\dot{H}$, for all $j\geq 1$. Resumming thus gives
\be
2\nabla^\alpha\nabla^\beta\left(\frac{\cG^{-1} -1}{\Box}\right)G_{\alpha\beta}\approx -\frac{9}{2}\left(\cG^{-1}(8H^2L^2)-1\right)\dot{H}\,.
\label{term3}
\ee
Finally, some more algebra reveals that the $\cI^{(i)}$'s in~(\ref{trace}) combine to give the following contribution at order $\dot{H}$:
\be
\cI^{(1)}+ \int_0^1 ds\left(\cI^{(2)}+\cI^{(3)}\right)\approx -\frac{27}{2}\l(8H^2L^2\r)^2\dot{H}\,.
\label{term4}
\ee

Combining~(\ref{term1})-(\ref{term4}), we obtain our modified cosmological equation
\be
0\approx 12H^2  + 6\dot{H} +  \frac{27}{4} \l(8H^2L^2\r)^2\dot{H}  + \frac{9}{2}\left(\cG^{-1}\l(8H^2L^2\r)-1\right)\dot{H}\,.
\label{later}
\ee
As a quick check, note that in the limit $HL\rightarrow 0$ we recover the standard vacuum equation: $R=0$.
However we are interested in the opposite limit, {\it i.e.}, $HL\gg 1$, in which the form factor
$\cG^{-1}$ is exponentially large, and~(\ref{later}) reduces to
\be
\frac{\dot{H}}{H^2} \approx -\frac{8}{3}\cG\l(8H^2 L^2\r)\,.
\label{eom2}
\ee
The right hand side is indeed $\ll 1$, confirming the validity of the slow-evolution approximation. This describes a self-inflating universe with effective slow-roll parameter
\be
\epsilon\equiv -\frac{\dot{H}}{H^2} \approx \frac{8}{3}\cG\l(8H^2L^2\r) \ll 1\,. 
\label{eps}
\ee
As mentioned earlier, this general definition of $\epsilon$ in terms of $\dot{H}$ reduces to the usual $\epsilon$ parameter in the case of scalar-driven inflation. 

\subsection{Perturbing Ricci} \label{altern1}

Our solution suggests an alternative derivation in terms of a perturbative expansion about de Sitter space:
\be
R_{\alpha\beta} = 3\bar{H}^2 g_{\alpha\beta}  + r_{\alpha\beta}\,,
\label{r}
\ee
where $\bar{H}$ is constant, and $r$ is a small correction characterizing deviations from de Sitter. To be precise, our approximation boils down to assuming a nearly constant Hubble parameter, which we can write as $H(t) = \bar{H} + h(t)$, with $h\ll \bar{H}$. The components of the full Ricci tensor at linear order in $h$ are then given by 
\bea
\nonumber
& & R^0_{\;\;0}= 3\left(H^2+\dot{H}\right)\approx 3\bar{H}^2+6\bar{H}h+3\dot{h}; \\
& & R^i_{\;\;j} = \left(3H^2+\dot{H}\right)\delta^i_{\;\;j}\approx \left(3\bar{H}^2+6\bar{H}h+\dot{h}\right)\delta^i_{\;\;j}\,,
\eea
from which we can read off the components of $r$:
\bea
\nonumber
& & r^0_{\;\;0} = 6\bar{H}h+3\dot{h} + \cO\left(h^2\right)\; \\
& & r^i_{\;\;j} = \left(6\bar{H}h+\dot{h}\right)\delta^i_{\;\;j} + \cO\left(h^2\right)\,.
\label{rcomp}
\eea

By adding total derivative terms to~(\ref{action1}), which therefore do not affect the variational principle, we can rewrite our action in the equivalent form
\bea
\nonumber
S &=& \frac{M_{\rm Pl}^2}{2}\int d^4x \sqrt{-g}\left\{ R + \l(G^{\alpha\beta}+3\bar{H}^2 g^{\alpha\beta}\r)\left(\frac{\cG^{-1}\l(\Box L^2\r) - 1}{\Box}\right)\l(R_{\alpha\beta}-3\bar{H}^2 g_{\alpha\beta}\r)\right\} \\
&=& \frac{M_{\rm Pl}^2}{2}\int d^4x \sqrt{-g} \left\{ R + \left(r^{\alpha\beta}-\frac{1}{2}g^{\alpha\beta}r \right)\left(\frac{\cG^{-1}\l(\Box L^2\r) - 1}{\Box}\right)r_{\alpha\beta}\right\} \,.
\label{action2}
\eea
Since $r$ is assumed small, it suffices to keep terms at most linear in $r$ when varying the action. After taking the trace, the equation of motion is approximately given by
\be
0\approx -12\bar{H}^2 - \cG^{-1}r + 2\nabla^\alpha\nabla^\beta\left(\frac{\cG^{-1} -1}{\Box}\right)r_{\alpha\beta}\,.
\label{trace2}
\ee
Substituting $r$ given in~(\ref{rcomp}) and, as before, neglecting terms of order $\dot{h}^2$, $\ddot{h}$, etc., we obtain
\be
\dot{h} \approx -\frac{8}{3}\bar{H}^2\cG\l(8\bar{H}^2 L^2\r)\,,
\label{eom3}
\ee
which is easily seen to agree with~(\ref{eom2}) by making the substitutions $\dot{h} = \dot{H}$ and $\bar{H}\rightarrow H(t)$.

\subsection{Newton's Version} \label{altern2}

A third method to obtain our solution, which will prove useful when calculating the spectrum of density perturbations,
starts out with a nearly de Sitter metric ansatz,
\be
ds^2 = - e^{2\phi(t)} dt^2 + e^{-2\phi(t)}  e^{2\bar{H}t}dx^2\,,
\label{newton}
\ee
where $\phi$ is assumed small. At linear order, $\exp(2\phi)\approx 1+2\phi$, and we recognize $\phi$ as a time-dependent Newtonian potential. 

Substituting into the first of Eqs.~(\ref{action2}) and truncating at quadratic order, we obtain the Lagrangian density
\bea
\nonumber
\frac{\cL}{M_{\rm Pl}^2} &= & 12\bar{H}^2\phi\l(1+\cO(\phi)\r)-3\phi\cG^{-1}\Box\phi-2\phi\left(\frac{\cG^{-1}-1}{\Box}\right)\left(\Box^2+18\bar{H}^2\Box+36\bar{H}^4\right)\phi \\
&+& 2\l(\nabla^\alpha\nabla^\beta\phi\r) \left(\frac{\cG^{-1}-1}{\Box}\right)\nabla_\alpha\nabla_\beta\phi + \cO(\phi^3)\,,
\label{quadphi}
\eea
where we have dropped an irrelevant constant term. In other words, we treat $\phi$ as a scalar field evolving in a background pure de Sitter
metric with Hubble constant $\bar{H}$. The resulting $\phi$ equation of motion is then
\bea
\nonumber
& & -\frac{3}{2}\cG^{-1}\Box\phi - \left(\frac{\cG^{-1}-1}{\Box}\right)\left(\Box^2+18\bar{H}^2\Box+36\bar{H}^4\right)\phi \\
& & \;\;\;\;\;\;\;\;\;\;\;\;\;\;\;\;\;\;\;\;\;\;\;\;\;+ \nabla^\alpha\nabla^\beta \left(\frac{\cG^{-1}-1}{\Box}\right)\nabla_\alpha\nabla_\beta\phi  =  -3\bar{H}^2\l(1+\cO(\phi)\r)\,.
\label{phieom}
\eea

Once again we make a ``slow-roll" approximation, keeping terms at most of order $\dot{\phi}$. In particular, this means we can neglect $\Box^2\phi$ and higher derivative terms. Moreover, based on our previous derivations we anticipate that this is a consistent approximation when $HL\gsim 1$, which we therefore assume from the outset. In this limit the dominant contribution from the left hand side comes from the last term:
\be
\nabla^\alpha\nabla^\beta \left(\frac{\cG^{-1}-1}{\Box}\right)\nabla_\alpha\nabla_\beta\phi \approx -\frac{9}{8}\cG^{-1}\l(8\bar{H}^2L^2\r)\bar{H}\dot{\phi}\,.
\label{phiapprox}
\ee
Thus, to leading order the equation for $\phi$ reduces to
\be
3\bar{H}\dot{\phi}\approx 8 \bar{H}^2\cG\l(8\bar{H}^2L^2\r)\,,
\label{phieom2}
\ee
which is reminiscent of the slow-roll evolution of the inflaton: $3H\dot{\varphi}\approx -V_{,\varphi}$. It will indeed be useful to think of the Newtonian potential as an inflaton in this context, especially when studying density perturbations. To make contact with our previous derivations, note that the Hubble parameter of Sec.~\ref{direct} is related through $\phi$ via time-reparametrization: $dt\rightarrow e^\phi dt$, and thus $\dot{H} \approx - H\dot{\phi}$. Substituting in~(\ref{phieom2}) and dropping the bars yields~(\ref{eom2}).

\section{Exiting the Self-Inflationary Phase and Reheating} \label{exit}

Since $\cG\l(8H^2L^2\r)$ is a decreasing function of $H$, while $\dot{H}$ is negative, eventually the rate of expansion reaches $|\dot{H}/H^2| \sim \cO(1)$, which signals the end of the self-inflationary phase. This occurs when $HL\sim \cO(1)$. 
At that point, the slow-evolution approximation breaks down, and one must revert to solving the full equation to describe the cosmological evolution. This daunting task requires numerical work which is beyond the scope of this paper. Nevertheless, if $\dot{H}$ remains negative throughout --- a mild assumption --- then within a few Hubble times the Hubble parameter will drop to $HL\ll 1$. In this regime, $\cG\rightarrow 1$, and~(\ref{action1}) reduces to Einstein gravity. Thus our assumption is that the universe makes a swift transition around $HL\sim \cO(1)$ from the self-inflationary phase to a decelerating phase described by the usual Friedmann equation.

Exiting inflation is of course not the end of the story --- the scenario must also account for reheating the universe. Since there is no inflaton to couple directly to matter fields in this case, the dominant reheating mechanism relies on gravitational particle production. As pointed out years ago by Ford~\cite{ford} and others~\cite{others}, the rapid change in the gravitational field at the end of inflation can excite matter fields which are not conformally coupled to gravity. (Particles of conformally invariant fields are not produced since the backround is conformally flat.) This results in a reheat temperature of order $H$ --- the Hawking temperature during inflation ---, which in our case is of order
\be
T_{\rm reheat} \sim L^{-1}\,.
\label{Treheat}
\ee
Of course this mechanism requires a sharp exit from inflation. A lingering accelerating phase would dilute the gravitationally produced particles, leaving the post-inflationary universe in an unacceptably cold state. This pitfall can be avoided in our case by choosing sufficiently steep form factors, such as our fiducial exponential example~(\ref{fidG}).

Nucleosynthesis constrains the reheat temperature to be at least 10~MeV, corresponding to an upper bound on the scale of new gravitational physics:
$L\lsim 100$~fm. We will later see that fixing the amplitude of density perturbations imposes a much tighter constraint  of $L\sim 10^{-22}$~cm.

\section{Validity of Effective Theory} \label{stability}

At first sight our action~(\ref{action1}) looks fine-tuned from the point of view of effective field theory: it is at most quadratic in Ricci, derivatives appear in a special exponential form, and so forth. Contrary to this expectation, in this section we argue that our self-inflating solution is in fact robust to a host of generic corrections to~(\ref{action1}). 

First consider adding terms to~(\ref{action1}) involving covariant derivatives acting on curvature tensors, such as
\be
R^{\alpha\beta}\nabla_\alpha\nabla_\beta R\,. 
\ee
(This is an interesting example because of its different tensor structure than the form factor in~(\ref{action1}).) Because of the derivatives, such terms vanish in the de Sitter limit and thus can {\it at best} be of order $\dot{H}$. Note that most terms will be higher order in $\dot{H}$ and thus negligible, such as $R^{\alpha\beta}\nabla_\alpha R\nabla_\beta R\sim H^4\dot{H}^2$. The above example gives
\be
R^{\alpha\beta}\nabla_\alpha\nabla_\beta R\approx -216H^4\dot{H}\,.
\ee
But even so, this is still negligible compared to the $\exp\l(64H^4L^4\r)\dot{H}$ contribution from the form factor. Thus generic corrections at order $\dot{H}$ have a much smaller coefficient than that of the exponential form factor. 

More worrisome are derivative-free terms, such as
\be
R^2L^2\sim H^4L^2\,.
\ee
Since $HL\gsim 1$ for our self-inflating solution, this is at least as important as the leading Einstein-Hilbert term. To see how it affects things, consider replacing the $H^2$ term in~(\ref{later}) with something of order $H^4L^2$, thereby modifying~(\ref{eom2}) to something of the form
\be
\frac{\dot{H}}{H^2} \sim H^2L^2\cG\l(8H^2 L^2\r)\,.
\ee
While the $H^2L^2$ factor makes the right hand side larger, nevertheless the exponential factor wins out and preserves the slow evolution approximation.

To summarize, corrections to the effective action that involve derivatives acting on curvature vanish in the de Sitter limit and thus are at best of order $\dot{H}$. Even so their coefficient is generically negligible compared to the exponentially large form factor. Derivative-free corrections are of the same order, or larger, than the Einstein-Hilbert term and thus cannot be neglected. However their influence in the equation of motion can effectively be absorbed in a ``renormalization" of the form factor $\cG$. Generically this renormalized form factor still vanishes in the UV, preserving the slow-roll approximation. One potentially important caveat to this analysis is whether the standard rules of effective field theory apply in this case. After all our original fading gravity action displays an infinite number of time derivatives and is therefore genuinely non-local. 

Be that as it may, a remarkable property of our effective action is that it becomes increasingly robust with increasing $HL$. That is, the higher the scale of inflation compared to the scale of new physics, the closer our solution is to pure de Sitter, and the less sensitive it becomes to generic corrections.

\section{Density Perturbations} \label{pert}

The growth of density fluctuations is most easily studied by perturbing~(\ref{newton}) in Newtonian gauge,
\be
ds^2 = - e^{2\phi(t)} \l(1+2\Phi(x,t)\r)dt^2 + e^{-2\phi(t)}\l(1-2\Phi(x,t)\r)e^{2\bar{H}t}dx^2\,.
\label{newton2}
\ee
Implicit in this ansatz is our neglecting anisotropic stress, which greatly simplifies the calculation. Although we have not checked this in detail, this approximation is likely consistent since we are restricting the analysis to vacuum equations of motion only. 

Postponing a careful analysis of this issue to future work, the virtue of neglecting anisotropic stress is that $\Phi(x,t)$ is now recognized as a perturbation in $\phi(t)$: $\phi\rightarrow \phi(t) + \Phi(x,t)$. Hence the quadratic lagrangian follows immediately from~(\ref{quadphi}),
\bea
\nonumber
\frac{\cL_{\rm quad}}{M_{\rm Pl}^2} &= & -3\Phi\cG^{-1}\Box\Phi-2\Phi\left(\frac{\cG^{-1}-1}{\Box}\right)\left(\Box^2+18\bar{H}^2\Box+36\bar{H}^4\right)\Phi \\
&+& 2\l(\nabla^\alpha\nabla^\beta\Phi\r) \left(\frac{\cG^{-1}-1}{\Box}\right)\nabla_\alpha\nabla_\beta\Phi + \bar{H}^2\cO\l(\Phi^2\r)\,,
\eea
where the precise coefficient of the last term is not important. The corresponding equation of motion is
\bea
\nonumber
& & -3\cG^{-1}\Box\Phi - 2\left(\frac{\cG^{-1}-1}{\Box}\right)\left(\Box^2+18\bar{H}^2\Box+36\bar{H}^4\right)\Phi \\
& & \;\;\;\;\;\;\;\;\;\;\;\;\;\;\;\;\;\;\;\;\;\;\;\;\;+ 2\nabla^\alpha\nabla^\beta \left(\frac{\cG^{-1}-1}{\Box}\right)\nabla_\alpha\nabla_\beta\Phi  =  \bar{H}^2\cO\l(\Phi\r)\,.
\label{Phieom}
\eea

In the short-wavelength regime, $k\gg aH$, we can neglect space-time curvature and set $\bar{H}$ to zero. Moreover, covariant derivatives commute in this limit. Thus the quadratic lagrangian
reduces to
\be
\cL_{\rm quad} \rightarrow -3M_{\rm Pl}^2\Phi\cG^{-1}\Box\Phi\,,
\ee
corresponding to $\cG^{-1}\l(\Box L^2\r)\Box\Phi_k = 0$, 
where $\Box$ now stands for the d'Alembertian in flat space. 
The canonically-normalized field variable, $\phi_c(x,t) = \sqrt{6}M_{\rm Pl}\cG^{-1/2}(\Box)\Phi$, therefore satisfies the usual Klein-Gordon equation. Since the form factor is unity on-shell, however, the adiabatic vacuum corresponds to the initial amplitude
\be
\Phi_k(t) \rightarrow \frac{1}{\sqrt{6}M_{\rm Pl}\sqrt{2k}}e^{ikt}\,.
\label{kgg1}
\ee

In the long-wavelength limit, $k\ll aH$, the behavior is analogous to that of $\phi(t)$ described above since $\Phi$ just represents a time-dependent shift of the background solution. Making a similar slow-roll approximation as in~(\ref{phiapprox}), we obtain $3\bar{H}\dot{\Phi}_k\sim \bar{H}^2\cG\l(8\bar{H}^2L^2\r)\Phi_k$, and thus
\be
\Phi_k(t) \approx {\rm const.}
\label{Phieom2}
\ee
for modes well outside the horizon.
 
Note that the short and long wavelength asymptotic solutions given above match those of inflationary scalar perturbations: the Newtonian potential oscillates as in flat space when the modes are well-inside the horizon, and asymptotes to a constant when the modes exit the horizon. Thus, taking into account the factor of $\sqrt{6}M_{\rm Pl}$ difference in normalization from a canonical scalar field, the $\Phi_k$ spectrum on super-horizon scales is given by
\be
\Delta\Phi_k \approx \frac{1}{\sqrt{12\pi}}\frac{H}{M_{\rm Pl}}\,.
\label{Phispec}
\ee
Alternatively we can derive~(\ref{Phispec}) using the horizon-crossing approximation, which corresponds to matching~(\ref{kgg1}) and the long-wavelength $\Phi_k\approx const.$ solution at $k=aH$. In standard inflationary calculations this is an accurate approximation whenever $H$ is slowly varying~\cite{wang}, which is the  case here. 

A useful gauge-invariant variable to track is the curvature perturbation on comoving slices, denoted as usual by $\zeta$~\cite{bardeen,mukhanov}, since it is
conserved on super-horizon scales, barring entropy perturbations~\cite{BST}. It is related to the Newtonian potential by
\be
\zeta = -\frac{H^2}{\dot{H}}\left(\Phi+\frac{\dot{\Phi}}{H}-\frac{\dot{H}}{H^2}\Phi\right)\,.
\ee
Since $\Phi\approx const.$ on large scales, and $\dot{H}/H^2\ll 1$, substitution of~(\ref{Phispec}) yields 
\be
\Delta \zeta_k \approx \frac{1}{\sqrt{12\pi}}\frac{H}{\epsilon M_{\rm Pl}} \,,
\label{zeta1}
\ee
where $\epsilon$ was introduced in~(\ref{eps}). 

As mentioned earlier, this differs by a factor of $\epsilon^{1/2}$ from the usual answer in scalar-driven inflation, tracing back to a difference in normalization of $\Phi$ in the two cases. This suggests that the tensor-to-scalar ratio is proportional to $\epsilon^2$ in this case, leading to a suppressed level of gravity waves. We will see in Sec.~\ref{GW}, however, that unlike inflation scale invariance for tensors is not generic in fading gravity models, although certainly possible. In particular, a sufficiently red tilt can easily overcome the $\epsilon^2$ suppression and yield a measurable level of gravitational waves on large scales.

\subsection{Fixing the Amplitude} \label{fixingamp}

\begin{figure}[ht]
\centering
\includegraphics[width=120mm]{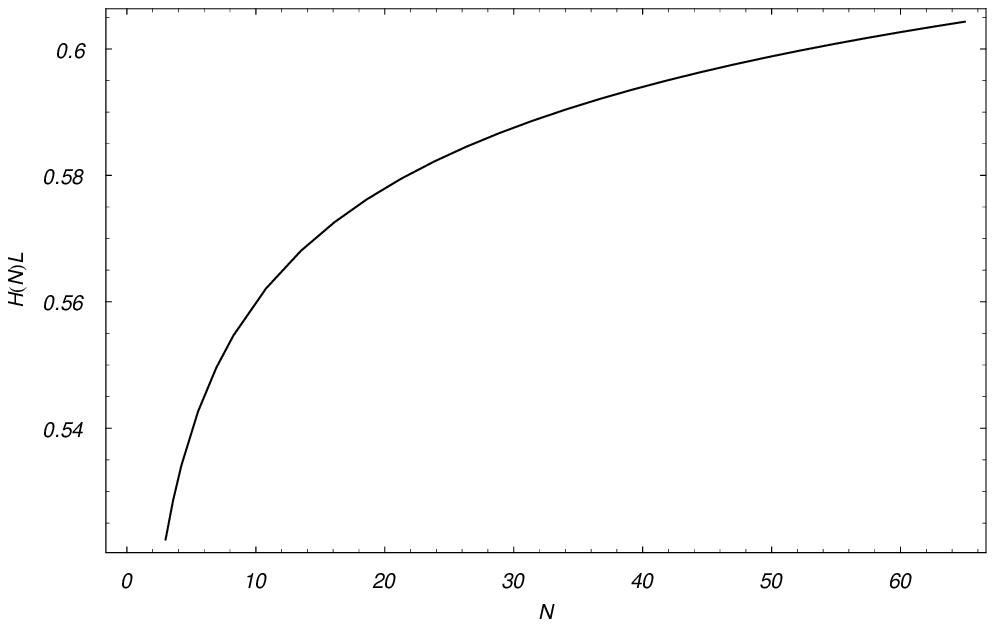}
\caption{Evolution of the Hubble parameter, $H(N)$, during self-inflation for $\cG=\exp\l(-\Box^2 L^4\r)$.}
\label{figH}
\end{figure}

\begin{figure}[ht]
\centering
\includegraphics[width=120mm]{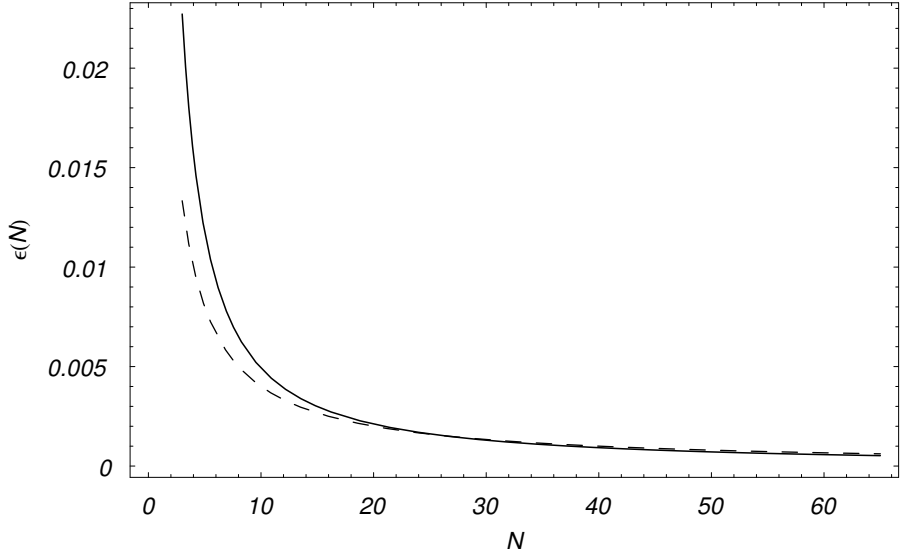}
\caption{Evolution of the effective slow-evolution parameter, $\epsilon (N) = - d\ln H/dN$, for the same choice of $\cG$ as before (solid line).
The fitting formula~(\ref{epsfit}), $\epsilon_{\rm fit}(N) = 1/25N$, is shown for comparison (dashed line).}
\label{figeps}
\end{figure}

In this section we match our expression for $\zeta$ with the observed large-scale amplitude of $10^{-5}$ from observations of the microwave background temperature anisotropy. To do so we need an expression for $H(N)$ and $\epsilon(N) \equiv - d\ln H/dN$, where $N$ is defined as the number of e-folds before the end of inflation:
\be
N\equiv \ln\left(\frac{a_{\rm end}H_{\rm end}}{aH}\right)\,.
\ee
For the fiducial form factor~(\ref{fidG}), the results of numerically integrating~(\ref{eom2}) are shown in Figs.~\ref{figH} and~\ref{figeps}. A useful fitting formula for the slow-evolution parameter is
\be
\epsilon_{\rm fit}(N) = \frac{1}{25N}\,,
\label{epsfit}
\ee
also shown in Fig.~\ref{figeps} for comparison. This agrees with the exact result to within 20-40\% except for the last few e-folds. Substituting in~(\ref{zeta1}) and using the fact that $H\sim L$ during inflation, the amplitude of perturbations on the largest scales must therefore satisfy
\be
\zeta \sim \frac{25N_{\rm obs}}{LM_{\rm Pl}}\sim 10^{-5}\,,
\label{zeta2}
\ee
where $N_{\rm obs}$ stands for the e-fold mark when the observable range of modes exits the horizon. 

All we need at this point is a relation between $N_{\rm obs}$ and $L$. The former can be expressed as
\be
N_{\rm obs}  = \ln\left(\frac{a_{\rm end}H_{\rm end}}{a_{\rm reheat} H_{\rm reheat}}\right) + \ln\left(\frac{a_{\rm reheat}H_{\rm reheat}}{a_{\rm eq} H_{\rm eq}}\right) + \ln\left(\frac{a_{\rm eq}H_{\rm eq}}{a_0 H_0}\right)\,,
\label{N1}
\ee
where ``eq" stands for matter-radiation equality, ``0" stands for today, and other subscripts are self-explanatory. 
The biggest unknown is the leading term since it depends on the detailed evolution between the end of self-inflation and reheating. For simplicity let us assume that the background evolves in this intervening phase as $a(t)\sim t^{1/3}$, as if dominated by the kinetic energy of a scalar field: 
\be
\ln\left(\frac{a_{\rm end}H_{\rm end}}{a_{\rm reheat} H_{\rm reheat}}\right) = \frac{2}{3}\ln\left(\frac{H_{\rm end}}{H_{\rm reheat}}\right)=-\frac{2}{3}\ln\left(\frac{T_{\rm reheat}}{M_{\rm Pl}}\right)\,.
\label{trans}
\ee
In the last step we have used the fact that reheating marks the onset of the radiation-dominated epoch, and thus $H_{\rm reheat} \sim T_{\rm reheat}^2/M_{\rm Pl}\sim H_{\rm end}^2/M_{\rm Pl}$. It is easy to check that the final result is insensitive to the details of the intervening phase --- it only changes the coefficient in~(\ref{trans}) by order one corrections.

Coming back to~(\ref{N1}), $z_{\rm eq} \approx 3000$ implies that the last term in~(\ref{N1}) is $\approx 4$. Moreover, since the universe is radiation-dominated between reheating and equality, we have $\ln(a_{\rm reheat}H_{\rm reheat}/a_{\rm eq}H_{\rm eq})\approx \ln (T_{\rm reheat}/T_{\rm eq})$. Putting everything together we obtain:
\be
N_{\rm obs}  \approx 4 + \ln\left(\frac{M_{\rm Pl}}{T_{\rm eq}}\right) + \frac{1}{3}\ln\left(\frac{T_{\rm reheat}}{M_{\rm Pl}}\right) \approx 65 - \frac{1}{3}\ln\left(L M_{\rm Pl}\right)\,,
\ee
where in the last step we have used $T_{\rm reheat}\sim L^{-1}$.

We can now substitute this expression for $N_{\rm obs}$ into~(\ref{zeta2}) and obtain a relation for $L$. Thus the observed large-scale amplitude of density perturbations fixes the scale of new physics to
\be
L \sim 10^{10}~{\rm GeV}\,,
\label{fixL}
\ee
which, unfortunately, lies well beyond the reach of laboratory tests of the gravitational inverse square law. Despite the relatively large number of e-folds required to explain the observed homogeneity and isotropy of our universe, the reheating temperature of $10^{10}$~GeV is moderately low, owing to the inefficiency of the reheating mechanism. It should be stressed that~(\ref{fixL}) applies to our exponential fiducial form factor --- more general form factors can yield higher reheating temperature.

\subsection{Scalar Spectral Index}

The power spectrum is given by $P(k)\sim\zeta^2\sim H^2/\epsilon^2$, where evaluation at horizon exit is understood. Since $\epsilon$ and $H$ are time-varying during inflation,
this leads to a spectral tilt defined as usual by
\be
n_s -1 \equiv \frac{d\ln P(k)}{d\ln k} = 2\left(\frac{d\ln H}{d\ln k}-\frac{d\ln \epsilon}{d\ln k}\right)\,.
\ee
The approximate correspondence between ``time" and ``scale" dependence is set as usual by the horizon-crossing condition, $k=aH$, which gives
$d\ln k\leftrightarrow dN$. 

For our fiducial choice of $\cG$, the equation of motion~(\ref{eps}) written in terms of $\epsilon$ implies
\be
\frac{d\ln \epsilon}{d\ln k} = - 4\epsilon\ln\left(\frac{3}{8}\epsilon\right)\,.
\ee
Moreover, since  $d\ln H/dN = -\epsilon$ by definition, the spectral index can be expressed as
\be
n_s -1 = 2\epsilon\left\{-1+4\ln\left(\frac{3}{8}\epsilon\right)\right\}\approx 8\epsilon\ln\left(\frac{3}{8}\epsilon\right)\,.
\ee
Numerically, we can substitute $N\approx 60$ in the fitting formula~(\ref{epsfit}) to obtain $\epsilon\sim 10^{-3}$, corresponding to $n_s\approx 0.96$. 
This is remarkably close to the best fit value from the WMAP 3-year data~\cite{wmap3}. Of course more general form factors will result in a spread of allowed $n_s$ --- this will be explored in subsequent work. However the indications are that the scalar spectrum in our model is degenerate with the generic single-field inflationary prediction~\cite{kst,boyle}: a red spectrum with a few percent deviation from scale invariance. Indeed, the scalar spectral tilt is unambiguously {\it red} if $\cG$ is monotonic. Since $\epsilon$ increases with time in this case, while $\dot{H}<0$, it follows that $H/\epsilon$ is a decreasing function of time, resulting in smaller amplitude for shorter wavelength modes.

\section{Primordial Gravitational Wave Spectrum} \label{GW}

The observational predictions of the model are so far qualitatively degenerate with the simplest models of scalar-driven inflation: nearly scale-invariant, adiabatic fluctuations, with a slightly red spectral tilt. A promising observable for distinguishability are tensor perturbations or primordial gravitational waves. 

We perturb the metric as
\be
g_{\alpha\beta} = \bar{g}_{\alpha\beta} + h_{\alpha\beta}\,,
\label{gw}
\ee
where for the purpose of this calculation the background metric $\bar{g}$ is taken to be exact de Sitter: $\bar{g}_{\alpha\beta}dx^\alpha dx^\beta =-dt^2 + e^{2Ht}dx^2$. The non-vanishing components for tensor perturbations are purely spatial, $h^0_{\;\;\alpha} = 0$, and satisfy the usual transverse, traceless conditions: $h^i_i=0$, $\partial_ih^i_j = 0$. The latter consist of four conditions on a symmetric tensor in 3 dimensions, leaving two degrees of freedom --- the two polarization states of the graviton. 

Our starting point is the action written in the form~(\ref{action2})
\be
S = \frac{M_{\rm Pl}^2}{2}\int d^4x \sqrt{-g} \left\{ R + \l(G^{\alpha\beta}+3H^2 g^{\alpha\beta}\r)\left(\frac{\cG^{-1} - 1}{\Box}\right)\l(R_{\alpha\beta}-3H^2 g_{\alpha\beta}\r)\right\}\,,
\label{nolambda}
\ee
where we have dropped the bars for simplicity. Substituting~(\ref{gw}) and dropping terms that are independent of $h$, we obtain a quadratic action for tensor modes
\be
S_{\rm tensor} =  \frac{M_{\rm Pl}^2}{2}\int d^4x  \sqrt{-\bar{g}} \frac{1}{4}h^i_{\;j}\l(\Box-8H^2\r)\left\{ 1+\left(\frac{\cG^{-1} - 1}{\Box}\right)\l(\Box-8H^2\r)\right\}h_{i}^{\;j}\,.
\label{pertds}
\ee
The constant piece in the kinetic term indicates that the usual graviton mode has a {\it positive} mass squared in this background. Already this is a significant difference from scalar-driven inflation where the graviton action describes two massless degrees of freedom.

\begin{figure}[ht]
\centering
\includegraphics[width=120mm]{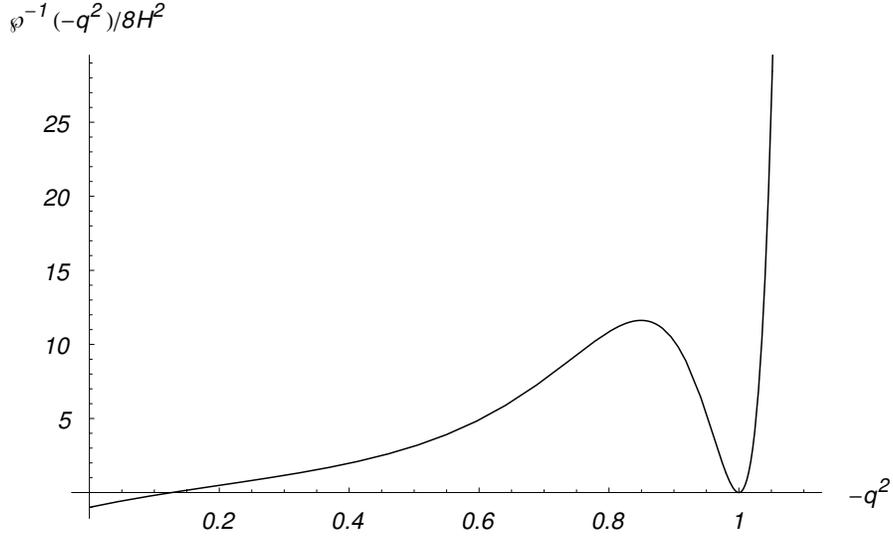}
\caption{Inverse graviton propagator in de Sitter background. Although difficult to distinguish in the plot, $\cP^{-1}(q^2)$ actually crosses zero at $-q^2\approx 1-\cG\l(8H^2L^2\r)$ and $-q^2=1$. Therefore the propagator has a total of 3 poles, with the one at $-q^2\approx 1-\cG\l(8H^2L^2\r)$ corresponding to a ghost.}
\label{propds}
\end{figure}

The above propagator is quite complicated and requires a careful analysis of its pole structure. To do so, consider a basis of eigenfunctions $h_{ij}^{(q)}$ for the $\Box$ operator, with dimensionless momentum eigenvalues $-q^2$:
\be
\Box h^{i\;\;(q)}_{\;j} = -8H^2q^2h^{i\;\;(q)}_{\;j} \,.
\label{q}
\ee
The inverse propagator, $\cP^{-1}$, is then given in terms of $q$ by
\be
\frac{\cP^{-1}(q^2)}{8H^2} = \l(-q^2-1\r)\left\{1+\left(\frac{\cG^{-1}\l(-8H^2L^2q^2\r) - 1}{-q^2}\right)\l(-q^2-1\r)\right\}\,,
\label{inverseprop}
\ee
which, as illustrated in Fig.~\ref{propds} for our fiducial form factor~(\ref{fidG}), has three zeros, corresponding to poles of the propagator. Although the figure focuses on a specific choice of $\cG$, we can argue that the number and locations of the poles are generic.

The zero at $-q^2=1$ is manifest from~(\ref{inverseprop}) and exists for any $\cG$. The other zeros can be found by setting the factor in curly brackets to zero to obtain
\be
1+q^2 - \cG\l(-8H^2L^2q^2\r)=0\,.
\label{otherzeros}
\ee
This is clearly satisfied near $-q^2=1$ at approximately $-q^2\approx 1-\cG\l(8H^2L^2\r)$. This, combined with the first pole, results in a quasi double pole near $-q^2=1$, which is a robust consequence of the approximation $\cG\l(8H^2L^2\r)\ll 1$ required for the background self-inflating solution. Finally, there is a third zero at small $-q^2$, denoted by $-q^2_{\rm small}$, which is easily seen by taking the log of~(\ref{otherzeros}):
\be
q^2_{\rm small} \approx \log\left( \cG(-8H^2L^2q^2_{\rm small})\right)\,.
\ee
For the fiducial form factor~(\ref{fidG}), this gives
\be
-q^2_{\rm small} \approx \frac{1}{64H^4L^4}\approx 0.12\,,
\ee
where in the last step we have used~(\ref{eps}) and~(\ref{epsfit}) with $N\approx 60$. The location of this third pole is in good agreement with the numerical result shown in Fig.~(\ref{propds}). While the double pole near $-q^2=1$ is robust under generic choice of $\cG$, we see that $-q^2_{\rm small}$ depends on the details of the form factor. Since it is precisely this pole which ends up determining the tensor spectral index, as we will see below, our model does not make a generic prediction of scale invariance for the gravitational wave spectrum.

The sign of the residue for the each of the poles can be read off from the slope of $\cP^{-1}$: positive slope corresponds to ``right sign" residue. Hence the pole at $-q^2\approx 1-\cG\l(8H^2L^2\r)$ is ghost-like. The  existence of ghost-like excitations in the momentum range where we trust the background solution is certainly worrisome, and we will devote Sec.~\ref{ghost} to analyze in detail the consequences of this instability.

While we have so far focused on helicity-2 modes of the metric, the modified propagator in~(\ref{pertds}) shows that our gravitons are massive and therefore are expected to have 5 polarizations. To be precise, the full quadratic lagrangian in the usual gauge, $\nabla^\alpha (h_{\alpha\beta} - \bar{g}_{\alpha\beta}h/2) = 0$, is given by
\be
{\cal L}_{\rm quad} \sim \frac{1}{4}\tilde{h}^{\alpha\beta}\l(\Box-8H^2\r)\left( 1+\left(\frac{\cG^{-1} - 1}{\Box}\right)\l(\Box-8H^2\r)\right)\tilde{h}_{\alpha\beta} -\frac{1}{16}h\Box h\,,
\label{pertds2}
\ee
where $\tilde{h}_{\alpha\beta} \equiv h_{\alpha\beta} - \bar{g}_{\alpha\beta}h/4$ is the traceless mode. Thus only the traceless part of the metric perturbation acquires a modified propagator, while the action for the trace part is unchanged compared to general relativity. In general we therefore expect 5 propagating degrees of freedom for each of the 3 poles, for a total of 16 (including the trace). Interpreting $-q^2$ as an effective mass from~(\ref{q}), each set of 5 polarizations should form a massive spin-2 representation. The analysis of Sec.~\ref{altern2} suggests that one of the corresponding helicity-0 states plays the role of an effective inflaton in our model, setting the clock for the cosmological evolution. A rigorous analysis of the propagating degrees of freedom in this model is currently in progress~\cite{future}.

\subsection{Tensor Spectrum}

Putting aside the ghost issue for the time being, let us charge ahead and compute the primordial gravitational wave spectrum. The calculation is closely related to its scalar counterpart. Well-within the horizon, the wave equation reduces to its flat space form, with usual Bunch-Davies solution
\be
h_k \rightarrow \frac{1}{M_{\rm Pl}a\sqrt{2k}}e^{ikt}\,.
\label{gwkgg1}
\ee
In the long wavelength limit, we have in principle three graviton excitations to keep track of, corresponding to the above three poles. Fortunately, however,
only the pole at small $-q^2$ is relevant --- the two poles near $-q^2=1$ correspond to massive gravitons and therefore
are not excited by the background expansion. The wave equation for this mode is approximately given by
\be
-\Box h^i_{\;j} = \ddot{h}^i_{\;j}+3H\dot{h}^i_{\;j}-2H^2h^i_{\;j}\approx 8H^2q^2_{\rm small} h^i_{\;j}\,,
\label{waveds}
\ee
which has the growing mode solution
\be
h_k = C_k a^{\frac{3}{2} \left( \sqrt{1+\frac{8}{9}\l(1+4q^2_{\rm small}\r)}-1\right)}\,.
\label{gwkll1}
\ee

The behavior at intermediate wavenumber requires solving the perturbation equation exactly, which is beyond the scope of this paper. For our purposes it suffices to use once again the horizon-crossing approximation, as we did for the scalar spectrum: we treat~(\ref{gwkgg1}) and~(\ref{gwkll1}) as exact solutions for $k>aH$ and $k<aH$, respectively, and determine $C_k$ by matching them at $k=aH$. The final answer for the large-scale gravitational wave spectrum is then
\be
\Delta h_k\equiv \frac{k^{3/2}}{\pi}|h_k|\sim \frac{H}{M_{\rm Pl}}\left(\frac{k}{H}\right)^{-\frac{3}{2} \left( \sqrt{1+\frac{8}{9}\l(1+4q^2_{\rm small}\r)}-1\right)}\,,
\label{hspec}
\ee
corresponding to a tensor spectral tilt
\be 
n_T = -\frac{3}{2} \left( \sqrt{1+\frac{8}{9}\l(1+4q^2_{\rm small}\r)}-1\right)
\label{nT}
\ee

Although apparently not generic, a nearly scale invariant tensor spectrum is possible if $-q^2_{\rm small}\approx 1/4$. It is straightforward to show that this
can be achieved with the following generalization of our fiducial form factor,
\be
\cG\l(\Box L^2\r) = \exp\left(\alpha\Box L^2 -\Box^2 L^4\right)\,,
\label{fidG2}
\ee
with $\alpha\sim\cO(1)$ suitably chosen. Even in this case, however, the amplitude of gravity waves is well below the sensitivity levels for near-future B-mode polarization experiments. Indeed the scalar amplitude~(\ref{zeta1}) implies a tensor-to-scalar ratio of $r\sim \epsilon^2$, which is at best $10^{-4}-10^{-3}$. 

The key implication of~(\ref{nT}), however, is that the tensor spectrum is generically not scale invariant --- its spectral tilt is sensitive to the choice of form factor. This is a distinguishing feature from scalar-driven inflation, where a nearly scale-invariant tensor spectrum follows directly from the slow evolution of the Hubble parameter
during inflation. In particular, a significant gravity wave amplitude on large scales requires a sufficiently red tilt, which is not hard to obtain. Our fiducial form factor, for instance, corresponds to $-q^2_{\rm small}\approx 0.1$, and thus $n_T\approx -1/3$, which is in fact too red to be consistent with observations. A more general form factor, such as~(\ref{fidG2}), can do the trick. 

The bottom line is that if gravitational waves are to be observed in our model, their spectrum must have a relatively large red tilt, which would distinguish it from standard inflation. Conversely, observing gravity waves with a nearly scale invariant spectrum {\it and} with a tensor-to-scalar ratio at the percent level would rule out fading gravity inflationary models. Of course failing to detect tensor perturbations would constrain but not rule out either standard or fading gravity inflation.

If futuristic CMB polarization or direct gravity wave experiments ever get down to sensitivity levels that would probe tensor-to-scalar ratios in the $10^{-4}-10^{-3}$ range, then our model would be strongly favored by the observation of a blue tilt for the tensor spectrum. This outcome, which is certainly allowed by~(\ref{nT}), would be an indisputable distinguishing feature from scalar driven inflation, where the spectrum is unequivocally red.

\subsection{Taming the Ghost} \label{ghost}

As usual it is possible to regularize the ghost instability by invoking Lorentz-violation at some high cutoff scale. We explore this possibility in this subsection.
In the limit $HL\gsim 1$, the ghost can be thought of as arising from an approximate double-pole at $-q^2 = 1$, or $\Box = 8H^2$. Indeed, in this regime the quadratic action~(\ref{pertds}) near $\Box = 8H^2$ is
\be
\cL_{\rm tensor} \sim \frac{1}{4}h^{ij}\cG^{-1}\l(8H^2L^2\r)\l(\Box-8H^2\r)^2h_{ij}\,.
\label{ghost1}
\ee
Incidentally, because of the $\cG^{-1}\l(8H^2L^2\r)$ prefactor the ghost decouples in the limit $HL\rightarrow \infty$. Unfortunately, as we will see in detail below, this decoupling limit leads to unacceptably large density fluctuations. The COBE constraint on the amplitude of $\delta\rho/\rho$ forces us to take the ghost by the horns.

\begin{figure}[ht]
\centering
\includegraphics[width=60mm]{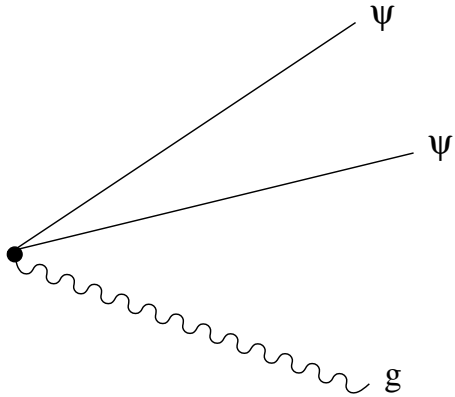}
\caption{Vacuum decay to ghost graviton and matter particles.}
\label{decay}
\end{figure}

A ghost indicates an explosive instability of the theory. Because ghost particles carry negative energy, the vacuum is unstable to decay, {\it e.g.}, through the process shown in Fig.~\ref{decay} in which a ghost graviton and two matter particles are spontaneously created from the vacuum. It is well known that the rate for this decay is formally infinite due to divergent phase space factors, even with a Lorentz invariant cut-off. The reason is simple: the energy carried by the ghost can be made arbitrarily large while keeping $p^2$ small, therefore satisfying some Lorentz invariant bound. 

The decay rate can be regularized by invoking explicit Lorentz breaking at some scale $\Lambda_{\rm LI}$ (measured in the cosmic rest frame) which sets an upper bound on the magnitude of the ghost energy~\cite{cline}. Ignoring irrelevant prefactors, the decay rate per unit volume is given by 
\be
\Gamma_{0\rightarrow g\psi\psi} \sim \frac{\Lambda_{\rm LI}^6}{M_{\rm Pl}^2}\cG\l(8H^2L^2\r)\,,
\ee
where the decoupling factor has trickled down from the propagator~(\ref{ghost1}).

Our self-inflating solution remains valid if the ghost energy density produced in a Hubble time,
\be
\Delta\rho_{\rm ghost} \sim \Lambda_{\rm LI} H^{-1}\Gamma_{0\rightarrow g\psi\psi}\,,
\ee
is less than the background effective energy density given by $3H^2M_{\rm Pl}^2$:
\be
\left(\frac{\Lambda_{\rm LI}}{M_{\rm Pl}}\right)^7\left(\frac{M_{\rm Pl}}{H}\right)^3\cG(8H^2L^2)\lsim 1\,.
\ee
This condition ensures that the rate of ghost production is sufficiently slow so as not to drown the inflating background. Self-inflation keeps going on in this case, continuously diluting the ghost and matter particles created spontaneously from the vacuum. 

It seems {\it a priori} that this bound can be trivially satisfied even for $\Lambda_{\rm LI}\sim M_{\rm Pl}$ by imposing a sufficiently high scale for inflation. However such a limit also makes the amplitude of perturbations too large. To see this explicitly let us substitute the expression for $\delta\rho/\rho$ using~(\ref{zeta1}) and~(\ref{eps}):
\be
\left(\frac{\Lambda_{\rm LI}}{M_{\rm Pl}}\right)^7\left(\frac{\delta\rho}{\rho}\right)^{-3}\cG^{-2}\l(8H^2L^2\r)\lsim 1\,.
\ee
Thus, unfortunately this has the decoupling factor with an inverse power. Substituting $\delta\rho/\rho\sim 10^{-5}$ and the value of $\cG\l(8H^2L^2\r)$ for the observable range of  modes --- see Sec.~\ref{fixingamp} --- yields a lower bound on the scale of Lorentz symmetry breaking:
\be
\Lambda_{\rm LI}\lsim 10^{15}\;{\rm GeV}\,.
\ee
Since COBE also constrains $L\sim 10^{10}$~GeV --- see~(\ref{fixL}) ---, this leaves us with a wide range of energy scales within which our effective description is valid.

Once inflation ends the ghost further decouples from all other fields. By construction at the perturbative level our theory has two propagating degrees of freedom about flat space, which are just the usual polarizations of the graviton. Thus in the limit $HL\rightarrow 0$ the ghost becomes non-perturbative and therefore decouples in weak field. This situation is in stark contrast with ghosts in other well-known higher curvature theories, such as $R+c_1R^2 + c_2 R_{\alpha\beta}R^{\alpha\beta}$~\cite{stelle}. In that case one generically finds an extra massive spin-2 field, which remains in the perturbative spectrum for arbitrarily large momenta. 

\section{Ghost Stories and the Ostrogradski Instability} \label{ostro}

The ghost instability described in the previous section is a specific manifestation of a general pathology afflicting higher derivative theories known as the Ostrogradski instability~\cite{ostro}. The basic statement is that the Hamiltonian of such theories generically depends linearly on all but one conjugate momentum variables and, hence, is unbounded from below. See~\cite{woodard} for a nice exposition. The result applies to theories with finite number of time derivatives, but will generically survive the infinite-derivative limit for theories that are analytic in the derivative operator. An example of the latter which will prove useful is
\be
\cL = \frac{1}{2}\phi \l(e^{\Box L^2}\Box+m^2\r)\phi\,.
\label{ostro1}
\ee

Such analytic form factors are ubiquitous in string theory. They appear in the truncated tachyon action in bosonic string field theory and in $p$-adic string theory. More generally such a structure is expected from the finite UV behavior. And indeed, it remains mysterious how string theory avoids the aforementioned instabilities~\cite{woodard}. The footprints of Ostrogradskian instabilities are found in time-dependent solutions of $p$-adic actions, where the scalar zooms by the tachyon vacuum and undergoes oscillations of ever-growing amplitude~\cite{barton}. In bosonic open string field theory, however, it has been argued recently that these are artifacts of the level truncation approximation, and that the full string field theory, albeit non-local in space, is local in lightcone time~\cite{gross}. See~\cite{gomis} for an analysis of degrees of freedom in $p$-adic and string field theory.

A related worry for higher-derivative theories is the lack of a well-defined Cauchy problem. To get a unique solution from an equation of motion with infinite number of time derivatives naively requires an infinite amount of initial data, which therefore amounts to arbitrarily specifying the solution over any finite time interval. 

These results afflict a wide class of non-local theories, including our model of fading gravity. However we would like to argue that the pathologies are of a mild form, at least in the regimes of interest for the cosmological analysis. Such considerations lead us to believe that in the neighborhood of the self-inflating solution there should exist an equivalent description of the theory which is local time and thus has a well-defined initial value formulation.

\subsection{Local Perturbative Locality}
 
An important exception to Ostrogradski's results is a non-local theory which can be made local by suitable invertible (non-local) field redefinitions. 
For instance, 
\be
\cL_0[\phi] = \frac{1}{2}\phi e^{\Box L^2}\Box \phi
\ee
is manifestly local in terms of a new field variable $\tilde{\phi} \equiv e^{\Box L^2/2}\phi$. However, adding local mass term and interactions,
\be
\cL[\phi] = \cL_0[\phi] + \sum_{n=2}^{\infty}g^n\phi^n\,,
\label{pertlocal}
\ee
makes the non-locality real and introduces Ostrogradskian pathologies. Nevertheless, as long as we restrict ourselves to perturbative analysis, with $\cL_0$ as the unperturbed Lagrangian, the theory has a well-defined initial value formulation:  all perturbative solutions only depend on the unperturbed solution and its time-derivative evaluated at some initial time. 
This can be made manifest by constructing an alternative theory which is local in time (albeit still non-local in space) and reproduces all perturbative solutions of the original theory. Such a construction is called {\it localization}. 

The point, of course, is that the runaway solutions of the full theory are genuinely non-perturbative. Perturbation theory effectively projects out these dangerous solutions and focuses on a subspace of solutions which is stable and spanned by two pieces of initial data. Theories of the form~(\ref{pertlocal}) are therefore said to be {\it perturbatively local}. While the perturbative regime circumvents the pathologies of the full theory, it is not completely worry-free since its solutions display acausality~\cite{woodard}, in the sense that corrections to the unperturbed solution at a given space-time point have support outside the past light cone. 

It is clear that the weak-field limit of our model about flat space satisfies perturbative locality:
\be
S_{\rm weak field} = \frac{M_{\rm Pl}^2}{8}\int d^4x \left\{\left(h^{\alpha\beta}-\frac{1}{2}\eta^{\alpha\beta}h\right)\cG^{-1}\l(\Box L^2\r)\Box h_{\alpha\beta}+\cO(h^3)\right\}\,.
\ee
In particular, one can make an invertible metric redefinition to render the kinetic term canonical, thereby pushing the non-local factors in the interaction terms. Even in more general backgrounds, the weak-field, flat space approximation is always valid within a sufficiently small region around any point. Thus our theory can be said to satisfy {\it local perturbative locality} --- the perturbative UV limit is free of the Ostrogradski instability and has a well-defined initial value formulation.

Beyond weak-field, however, the perturbed action about general backgrounds is not generically perturbatively local. 
Consider for instance our quadratic action~(\ref{pertds}) for transverse, traceless perturbations about the self-inflating solution. Because of the extra poles in the propagator, in this case the non-locality cannot be eliminated through an invertible field redefinition. This is akin to~(\ref{ostro1}). The Ostrogradski instability is signaled by the existence of the ghost-like mode discussed in Sec.~\ref{GW}. 

\subsection{Non-Perturbative de Sitter Solution}

To summarize, our toy model action~(\ref{action1}) for weaker gravity displays local perturbative locality but suffers from the non-perturbative Ostrogradskian instability. Moreover the action is non-local in time and is therefore bereft of a Cauchy problem. The non-perturbative nature of the self-inflating solution makes one wonder whether it secretly exploits unwanted instabilities of the theory. A related worry pertaining to the temporal non-locality of the full theory is whether the de Sitter solution requires fine-tuning an infinite number of time derivatives, and whether we could have instead obtained any other background evolution.

While these questions are certainly warranted, it is encouraging to note that the self-inflating solution appears to rely in the most minimal way possible on the higher-derivative nature of the theory. Indeed, despite having an infinite number of time derivatives, the modified cosmological equation~(\ref{eom2}) is nevertheless a second-order differential equation in the scale factor, just as in Einstein gravity. This suggests that it should be possible to construct an alternative theory which would be local in time (although likely non-local in space) and would reproduce all dynamics within a solution-space neighborhood of the self-accelerating solution. 

These considerations naturally lead us to ask whether our self-inflating scenario is degenerate with some (local) scalar field model. While we cannot completely rule out this possibility, the allowed strong departure from scale invariance in the tensor spectrum suggests that such a scalar-field analogue is impossible. Our gravity wave spectrum can be blue. In Einstein gravity, on the other hand, a blue tilt for tensors requires either a contracting universe (as in ekpyrotic/cyclic models) or an expanding universe driven by a fluid which violates the null energy condition. 

The former is unlikely since contracting universes generically lead to a big crunch singularity; although we haven't ruled out a singularity at the end of the self-inflating phase, everything so far points towards a graceful exit and smooth matching onto a radiation-dominated era. Degeneracy with null-energy violating models is also unlikely. First of all, the only consistent candidate known to this author is the recent proposal of~\cite{paolo} based on ghost condensation~\cite{nimaghost}. In this case, consistency of the model requires $\dot{H}\ll H^2$, which yields only a mild blue tilt. Even if one could extend the validity of the model such that a strong blue tilt obtains, this would likely translate into large deviations from scale invariance in the scalar spectrum as well. Degeneracy with local scalar field models is therefore unlikely. If a temporally-local alternative description of the self-inflating solution is possible, it will likely retain the spatial non-locality.

\section{Conclusions and Future Directions} \label{conclu}

In this paper we have studied the cosmology of a toy model where the gravitational force becomes exponentially weak at short distances.
This is accomplished by modifying the graviton propagator by an analytic form factor which shuts off in the UV. In this context we found novel inflationary
solutions that do not rely explicitly on scalar fields or other form of stress energy. Instead, inflation results from the modified vacuum equations of motion. 
Of course there are new degrees of freedom associated with the modified propagator, in particular a scalar mode which acts as a clock for the inflationary evolution.
What is intriguing, however, is that this effective inflaton effectively decouples and therefore disappears from the perturbative spectrum in the flat space limit. 

A novel inflationary mechanism is interesting because it opens the door to a new realm of  ``generic" predictions. Moreover, what may appear as fine-tuning in
scalar-driven inflation could be more natural in the fading gravity context, and vice versa.

A remarkable feature is that de Sitter space is an increasingly accurate solution to the fading gravity equations as $HL\rightarrow\infty$. Generic corrections to the effective action become less and less relevant for the background solution in this limit. Usually the exact opposite happens: corrections to the effective action become more important above the scale of new physics. The stability of our solution originates from the special property that the Ricci curvature tensor for pure de Sitter vanishes when acted upon by covariant derivative operators. Thus, pure de Sitter is oblivious to the form factor, but small deviations from it are highly sensitive to its action in the limit of large $HL$.

This work was mostly exploratory in nature, and many important issues must be resolved for the scenario to be viable:

\begin{itemize}

\item The details of the transition from self-inflation to a radiation-dominated universe must be worked out in detail. There are compelling
reasons to believe such a transition is possible: self-inflationary dynamics occur with $|\dot{H}|/H^2$ increasing, so that the slow-evolution approximation eventually breaks down;
radiation is produced through particle production; as $HL\rightarrow 0$ the theory reduces to Einstein gravity. While these are compelling hints, it is nevertheless 
conceivable that the self-inflationary phase and the normal radiation-dominated evolution actually lie on different branches of solutions or are separated by a curvature singularity. This is currently under investigation~\cite{future}.

\item The perturbation analysis around the self-inflating solution reveals a massive tensor mode which is ghost-like, indicating a violent instability. We have used standard arguments invoking Lorentz symmetry breaking at high energy to regularize the instability and render our solution viable. A better understanding of the origin of this ghost might 
suggest an improved, ghost-free version of the model.

\item A distinguishing prediction is that the spectrum of gravity waves is generically not scale invariant. In particular a relatively strong red tilt is possible, without jeopardizing the scale invariance of the scalar spectrum. It would be interesting to study in greater detail the observational constraints, as well as the implications for microwave polarization experiments and gravitational wave detectors.

\item From a phenomenological perspective, other potential observable signatures should be investigated, such as the predicted level of non-gaussianity. 

\end{itemize}

\bigskip
\noindent
{\bf Acknowledgments}

I am grateful to Alberto Nicolis for many insightful discussions and collaboration in the early stages. I would like to thank A.~Aguirre, N.~Arkani-Hamed, T.~Banks, C.~Burgess, P.~Creminelli, S.~Dubovsky, L.~Ford, J.~Gomis, R.~Jaffe, R.~Woodard, B.~Zwiebach, and especially C.~de~Rham and A.~Tolley for helpful discussions.

\end{document}